  \documentclass[a4paper,fleqn,usenatbib,onecolumn]{mnras}
\usepackage[T1]{fontenc}
\usepackage{ae,aecompl}
\usepackage{graphicx}
\usepackage{amsmath}
\usepackage{amssymb}
\newcommand{\VEC}[1]{\vec {#1} }
\title[
Evolution of magnetic deformation in neutron star crust
]{
Evolution of magnetic deformation in neutron star crust
}
\author[Y. Kojima, S. Kisaka, K. Fujisawa]{Yasufumi Kojima$^1$, 
\thanks{%
E-mail: ykojima-phys@hiroshima-u.ac.jp}
Shota Kisaka$^{1,2,3}$, 
Kotaro Fujisawa$^4$\\ 
$^1${Department of Physics, Hiroshima University, Higashi-Hiroshima, Hiroshima 739-8526, Japan}\\
$^2${Frontier Research Institute for Interdisciplinary Sciences, Tohoku University, Sendai, 980-8578, Japan}\\
$^3${Astronomical Institute, Graduate School of Science, Tohoku University, Sendai, 980-8578, Japan}\\
$^4${Research Center for the Early Universe, Graduate School of Science, University of Tokyo, Bunkyo-ku, Tokyo 113-0033, Japan}
}
\begin{document}
  \maketitle
\begin{abstract}
   In this study, we examine the magnetic field evolution occurring in a neutron star crust.
Beyond the elastic limit, the lattice ions are assumed to act as a plastic flow.
The Ohmic dissipation, Hall drift, and bulk fluid velocity driven 
by the Lorentz force are considered in our numerical simulation.
A magnetically induced quadrupole deformation is observed in the crust during the evolution. Generally, the ellipticity decreases as the magnetic energy decreases.
In a toroidal-field-dominated model, the sign of the ellipticity changes. 
Namely, the initial prolate shape tends to become oblate.
This occurs because the toroidal component decays rapidly
on a smaller timescale than the poloidal dipole component. 
We find that the magnetic dipole component does not change significantly 
on the Hall timescale of $\sim 1$Myr for the considered simple initial models.
Thus, a more complex initial model is required to study the fast decay of surface dipoles on the abovementioned timescale.
\end{abstract}


\section{Introduction}
The magnetic field strength of a single neutron star ranges 
from $10^{8}$ to $10^{15}$ G
\citep{2015RPPh...78k6901T,2019RPPh...82j6901E}.
Neutron stars exhibit different phenomena, based on which they are broadly classified.
The diversity of the classes could be simply attributed to their formation and environment. 
It is also known that 
the activity of magnetars, a class of strongly magnetized stars, originates from their magnetic energy.
Therefore, the magnetic field is not constant in some neutron stars of ages
$\sim 10^{4}$--$10^{5}$yr.
In a theoretical study, it is important to understand the extent of the
magnetic field change on the abovementioned timescale.
The physics of the magnetic field evolution in neutron stars is discussed in a classical 
paper\citep{1992ApJ...395..250G}.
The evolution is determined by the advection and 
dissipation of the magnetic field.
Typically, numerical simulations are required 
because of the nonlinear nature of the coupled equations.
In a neutron star core, the Ohmic dissipation is less effective owing 
to the high conductivity.
For a fast evolution on the timescale of $<10^{6}$yr, the transportation of the magnetic flux 
to the lower conductivity region in the crust or near the surface is important.
The core of a neutron star comprises multi-components; it is primarily a fluid mixture of 
neutrons, protons, and electrons, whose number densities vary 
from the center to the crust.
In a weak-coupling regime, they can move with different velocities 
so that the transport velocity is multiple.
In addition, the effects of the superfluidity and superconductivity 
in the core increase the complexity of the relevant physics.
Although this challenging issue has been addressed in the past
\citep[e.g.,][]{2011MNRAS.413.2021G,2015MNRAS.453..671G,
2016MNRAS.456.4461E,2016MNRAS.462.1453G,
2017MNRAS.469.4979P,2017MNRAS.471..507C,2018MNRAS.473.2771B},
it is still being debated\citep{2017PhRvD..96j3012G,
2018MNRAS.473.4272K,2018PhRvD..98d3007O,2019MNRAS.485.4936G,
2020MNRAS.tmp.2460C}.
The physics of the crustal magnetic field is comparatively simple.
In the solid crust, the lattice ions are fixed, 
and only the electrons are mobile.
The magnetic field evolution due to the Hall drift 
and the Ohmic dissipation has been investigated by several researchers.
A magnetic field with an axial symmetry was numerically simulated 
on a secular timescale
\citep[e.g.,][]
{2007A&A...470..303P,2012MNRAS.421.2722K,2013MNRAS.434..123V}.
Evolution to an equilibrium configuration and its stability  
were was discussed in
\citep{2013MNRAS.434.2480G,
2014PhRvL.112q1101G,2014MNRAS.438.1618G,2014ApJ...796...94M}.
Recently, the numerical simulation of the long-term magnetic-field 
evolution has been extended to three-dimensional models
\citep{2015PhRvL.114s1101W,2016PNAS..113.3944G,
2018ApJ...852...21G,2019CoPhC.237..168V},
revealing some of the effects ignored in the two-dimensional models.
In this study, we explore the effects of the plastic flow in the crust of a neutron star.
In some strongly magnetized stars, the field strength is extremely high, 
deforming the lattice structure of the ions by the shear stress.
A crust responds elastically when its deformation
is within a critical limit, beyond which it cracks or responds plastically.
The breaking strain determined by molecular dynamics 
simulations\citep{2009PhRvL.102s1102H} is $\sim 0.1$, whereas 
semi-analytical methods
\citep{2010MNRAS.407L..54C,2018MNRAS.480.5511B}
give a smaller value $<0.04$.
Abrupt crust breaking can cause a magnetar outburst and/or a rapid 
radio burst\citep{2015MNRAS.449.2047L,2016ApJ...833..189L,
2019MNRAS.488.5887S,2020PhRvD.101h3002S,2020arXiv200801114I}.
Thus, understanding the response of the crust of a neutron star is crucial for improving its 
astrophysical modelling.
The plastic flow beyond the critical limit was
explicitly demonstrated by a long-term simulation of a two-dimensional
box\citep{2019MNRAS.486.4130L}, and
the effect of the large-scale velocity circulation occurring in an axially symmetric crust
was explored \citep[][hereafter referred as Paper I]{2020MNRAS.494.3790K}. 
The resultant bulk velocity was
approximately proportional to the square of the magnetic field strength,
whereas the Hall velocity varied linearly with the strength.
Therefore, in magnetars or neutron stars with a hidden 
strong field, the effect of the plastic flow 
on the crustal field evolution is important.
We assume that the plastic flow occurs at all times in the entire crust.
Actual neutron star crust might be a mixture of the elastic/plastic responding regions, since magnetic stress depends on the place, and exceeds a threshold somewhere. 
The flow velocity $v_{b}$ therefore arises not in the entire crust but in super-critical region only.
For such a partially flowing structure, we have to solve an elliptic equation explained in next section with the constrain $v_{b}=0$ in the elastic region.
The numerical method becomes more complicated, and the solution may contain local irregularities.
Our concern is a global aspect so that we compare a flowing model ( $v_{b}\ne 0$) in the entire crust with no flowing model, i.e. the Hall evolution model.
Realistic model probably corresponds to the intermediate between these limiting cases.
In this study, we also investigate the effect of the strong magnetic field 
in a neutron star crust.
It is well known that strong magnetic fields deform stellar shapes.
We numerically simulate the time evolution of a quadrupole deformation. 
This problem was previously studied by \citet{2016MNRAS.459.3407S}, who claimed that large deformations can occur in normal radio pulsars
having a surface dipole field of $10^{12}$--$10^{13}$G.
Some neutron stars experience a large quadrupole deformation owing to a hidden field, which is unrelated to the inferred dipole strength.
This topic is extremely important in gravitational wave astronomy
\citep[e.g.][]{2017ApJ...844..112G,2019MNRAS.490.2692K}.
Namely, potential sources for continuous gravitational waves in near future detection are revealed.
Our approach to calculate the magnetic deformation is different from that in the referred study; therefore, it is important to compare the results from both the methods.
The models and the equations of this study are presented in Section 2. In Section 3, 
the numerical results of the magnetic energy and the surface dipole moment 
are discussed.
In Section 4, the quadrupole deformation of a neutron star crust is described.
Finally, our concluding remarks are presented in Section 5.

\section{Mathematical formulation}
When a magnetized neutron star is born, all the forces are 
in equilibrium on a short dynamical timescale.
The magneto-hydrodynamic(MHD) equilibrium 
among the pressure, gravity, and Lorentz force is expressed as
\begin{equation}
-{\VEC \nabla}P-\rho{\VEC \nabla}\Phi_{\rm G}
+\delta {\VEC f} =0,
  \label{Forcebalance.eqn}
\end{equation}
where magnitude of the last term,
$\delta {\VEC f} \equiv c^{-1}{\VEC j}\times {\VEC  B} $, 
is $\sim 10^{-7}(B/10^{14}{\rm G})^2$ times smaller than
those of the first and second terms.
We ignore the stellar rotation.
Any deviation from the static shape with spherical symmetry 
is sufficiently small to 
be considered as a perturbation in the background equilibrium.
Equation (\ref{Forcebalance.eqn}) expressing a spherical equilibrium is reduced to
\begin{equation}
 \frac{dP}{dr}=-\frac{G\rho M_{r}}{r^2},
 \label{eqnbackground}
\end{equation}
where $M_{r}$ is the mass contained within a sphere of radius $r$.
This study is limited to an axially symmetric magnetic field configuration.
The poloidal and toroidal components of the magnetic field are expressed as
\begin{equation}
\VEC{B}=\VEC{\nabla}\times \left(
\frac{\Psi}{\varpi}\VEC{e}_{\hat{\phi}}\right)
+\frac{S}{\varpi}\VEC{e}_{\phi} ,
\label{eqnDefBB}
\end{equation}
where $\varpi=r\sin\theta$ is the cylindrical radius.
Under barotropic MHD equilibrium, 
the current function, $S$, should be a function of 
$\Psi$, and the azimuthal current is expressed as 
\citep{2005MNRAS.359.1117T,2013MNRAS.434.2480G,
2014MNRAS.445.2777F}
\begin{equation}
\frac{4\pi\varpi}{c}j_{\phi} = -\rho K^\prime \varpi^2 
 -S^\prime S.
 \label{MHDeqil.eqn}
\end{equation}
In Eq.(\ref{MHDeqil.eqn}),
$K$ is also a function of $\Psi$, and  
a prime denotes a derivative with respect to $\Psi$.
The Lorentz force is defined as
\begin{equation}
\delta {\VEC f}=\rho K^{\prime}{\VEC \nabla}\Psi.
\end{equation}
For given functional forms $S(\Psi)$ and $K(\Psi)$, 
the magnetic function, $\Psi$,
which describes the poloidal field, ${\VEC B}_{\rm p}$, by eq.(\ref{eqnDefBB}),
is solved for the source term (\ref{MHDeqil.eqn}).
Although this method is simple, the actual calculations are difficult.
It is not possible to obtain the solution 
for arbitrary forms of $S(\Psi)$ and $K(\Psi)$.
Presumably, the barotropic condition highly constrains static models.
In an MHD equilibrium, electrons are not in equilibrium; therefore, the magnetic field tends to a configuration determined by the Hall equilibrium on a secular timescale
\citep{2013MNRAS.434.2480G,2014MNRAS.438.1618G}.
The Lorentz force, ${\VEC j}\times {\VEC B}$, 
is not fixed but changes according to the magnetic field evolution, which is governed by the induction equation,
\begin{align}
\frac{\partial}{\partial t}{\VEC B}=-{\VEC \nabla}\times (c{\VEC E}),
    \label{Frad.eqn}
\end{align}
where
\begin{equation}
c{\VEC E}=\frac{c}{\sigma}{\VEC j} 
+\frac{1}{e n_{\rm e}}{\VEC j}\times {\VEC B}
-{\VEC v}_{\rm b}\times {\VEC B}.
    \label{Edef.eqn}
\end{equation}
In eq.(\ref{Edef.eqn}), $\sigma$ is the electric conductivity, 
$ n_{e}$ the electron number density, and 
${\VEC v}_{\rm b}$ the bulk flow velocity.
The latter may be approximated as the ion velocity owing to
the large mass difference between ions and electrons. 
In a comoving frame of a neutron star, the ions in the crust are fixed 
in the lattice, i.e., ${\VEC v}_{\rm b}=0$.
In this case, the dynamics expressed in eq.(\ref{Frad.eqn})
is determined by the magnetic field because
of its relation to the electric current, ${\VEC j}$, by Amp{\'e}re--Bio-Savart's law as follows:
\begin{equation}
  {\VEC j}=\frac{c}{4\pi}{\VEC \nabla}\times {\VEC B}.
    \label{Amp.eqn}
 \end{equation}
The magnetic field evolution changes the structure and disrupts the abovementioned force equilibrium. 
This process may be slow or abrupt, depending on the material property. 
When a magnetic stress exceeds a threshold, the crust 
may be fractured, and the magnetic field is rearranged. 
Abrupt crust-quakes can cause a magnetar outburst and/or a rapid radio burst.
In this study, we consider the opposite case.
Specifically, we treat the gradual deforming motion of the lattice ions as a plastic flow, whose velocity is expressed by ${\VEC v}_{\rm b}$.
The effect of the magnetic field evolution is added as
a new velocity component ${\VEC v}_{\rm b}$ in eq.(\ref{Edef.eqn}).
The method to determine ${\VEC v}_{\rm b}$ is described in Paper I
\citep{2020MNRAS.494.3790K}.
Here, we briefly explain the basic concept of the proposed procedure.
The Lorentz force, $\delta {\VEC f}(\equiv c^{-1}{\VEC j}\times {\VEC  B})$,
is generally decomposed as a sum of irrotational and solenoidal vectors as follows:
\begin{equation}
\delta {\VEC f} =-{\VEC{\nabla}} \delta \chi 
+{\VEC{\nabla}}\times \left(\delta {\VEC A}\right).
\label{eqndecomp}
\end{equation}
The scalar potential, $\delta \chi$, and the vector potential, 
$\delta {\VEC A}$, satisfy
\begin{align}
\nabla^2 \delta \chi =-\VEC{\nabla}\cdot \delta {\VEC f},
\label{eqndf2}
\\
{\VEC{\nabla}}\times {\VEC{\nabla}}\times \delta {\VEC A}
= \VEC{\nabla}\times\delta {\VEC{f}}.
 \label{eqncirc}
\end{align}
The potential, $\delta \chi$, is closely coupled to the 
pressure and gravity terms in eq.(\ref{Forcebalance.eqn}).
However, the solenoidal part of $\delta {\VEC f}$ 
is almost irrelevant for these dominant forces. 
It drives a shear flow and leads to a nonbarotropic distribution, 
$\VEC{\nabla}(P+\delta P)\times \VEC{\nabla}(\rho +\delta\rho)\neq 0$.
The force in a strongly magnetized star may deform its crust plastically.
This effect is incorporated as an extra viscous flow term in 
the equation of motion, and the dynamics is expressed by the 
standard Navier--Stokes equation\citep{2016ApJ...824L..21L,2019MNRAS.486.4130L}.
The solenoidal part of the Lorentz force is assumed to be
balanced by the shear stress of the viscous bulk motion (see Paper I)
An incompressible fluid approximation is also assumed for simplicity.
Because our focus is on the long-term behavior of the magnetic field,
all the forces are always approximated to be instantaneously balanced.
Therefore, the velocity field structure is 
determined under a stationary condition.
The balance between the viscous fluid motion and the Lorentz force suggests that the bulk velocity field, $v_{\rm b}$, depends on 
$\propto \nu^{-1}B^2$, where $\nu$ is the viscous coefficient and $B$ is the typical magnetic field strength.
The Hall velocity, $v_{\rm H}\equiv |j/(e n_{\rm e})|$, is proportional to $B$.
Eq.(\ref{Edef.eqn}) contains two advection velocities; however, the bulk velocity, $v_{\rm b}$,
is important in low-viscosity and strong-magnetic-field regimes.
The former condition is generally satisfied near the stellar surface.
We found that the typical value of $\nu=$ 
$10^{36}-10^{37}{\rm{{g}{cm}}}^{-1}{\rm{s}}^{-1}$ is important for a magnetized neutron star with $B_0\sim 5\times 10^{13}$G at the pole
\citep[][Paper I]{2019MNRAS.486.4130L}.
Herein, the behavior on the smaller timescale ($\propto v_{\rm b}^{-1}$) 
associated with bulk motion is explored.
This study also examines the 
 effect of the time variation of the irrotational part of the Lorentz force.
The potential, $\delta \chi$, is important for magnetic deformation 
because it may be regarded as
an additional pressure term in eq.(\ref{Forcebalance.eqn}).
Its effect on the magnetic field evolution is less clear because 
it strongly depends on the responses of the dominant forces.
Herein, we ignore this, and 
only in Section 4 present the study of the magnetic deformation 
using evolutionary models free from the irrotational part.

\section{Evolution of magnetic energy}
In this section, we consider the effect of the bulk fluid motion on the magnetic evolution. 
Specifically, the magnetic field evolution of a purely Hall evolution is determined by eqs.
(\ref{Frad.eqn}),(\ref{Edef.eqn}), and (\ref{Amp.eqn}) with
${\VEC v}_{\rm b}=0$.
The effect of the bulk motion is incorporated in eq.~(\ref{Edef.eqn})
by including ${\VEC v}_{\rm b}$, 
which is determined by the solenoidal part of the Lorentz force.
Two types of evolutionary models are compared.
%

   \subsection{Barotropic MHD equilibrium}
For the initial magnetic field configuration, a reasonable assumption is barotropic MHD equilibrium, 
for which the azimuthal current is expressed in eq.(\ref{MHDeqil.eqn}).
The solution for a purely poloidal field ($S=0$) with a constant 
$K^\prime$ is easily obtained.
We denote the solution as $\Psi_{(1)}$ and the current as 
$j_{\phi (1)}\equiv -c\rho K^\prime\varpi/4\pi$. 
For this case, the angular dependence of the magnetic field is dipolar,
i.e., $\Psi_{(1)}~(\propto \sin^2\theta)$ is expressed by a single 
Legendre polynomial of $l=1$.
During the magnetic evolution, 
a toroidal magnetic field is generally induced, 
and the symmetry with respect to the angular dependence 
is not conserved owing to the nonlinear coupling terms in eqs.
(\ref{Frad.eqn})--(\ref{Edef.eqn}).
We first discuss the magnetic field evolution from the 
initial state, $\Psi_{(1)}$, in the crust.
This model is referred as the fiducial model.
The initial magnetic field and current distribution were demonstrated in 
Fig.1 of Paper I
\footnote{
Animation of time-variation of the Lorentz force 
is available in the online version. 
Three components during the Hall evolution of the fiducial model
are shown by color contour.
}.

\begin{figure}
\begin{center}
  \includegraphics[scale=0.80]{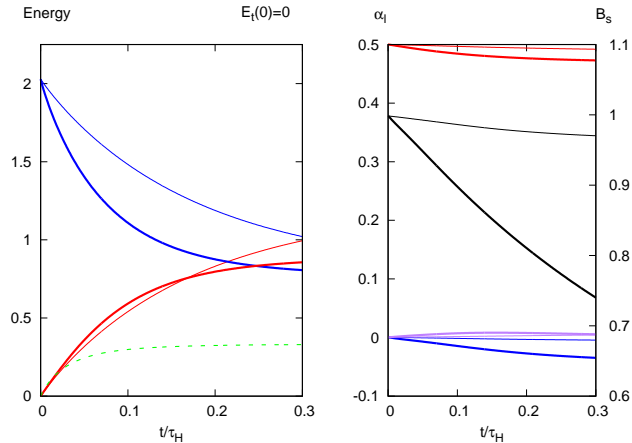}%
\caption{ 
 \label{Fig1}
Left panel shows decay of magnetic energy.
Magnetic energy $E_{\rm B}/(B_{0}^2R^3)$ is plotted as decreasing curve
(in blue color), time-integrated Joule loss as increasing curve (in red color),
and work to bulk motion as dotted curve (in green color).
Right panel displays time-variation in magnetic multipole moments,
$\alpha_{1}/(B_{0}R^{3})$ (in red color),
$\alpha_{3}/(B_{0}R^{5})$ (in blue color),
$\alpha_{5}/(B_{0}R^{7})$ (in purple color).
Field strength 
$B_{r}/B_{0}$ at polar surface with right scale 
is plotted by black curve.
Thick curves represent model in which bulk motion is considered.
}
\end{center}
\end{figure}

%
Below, we briefly discuss the model.
Our consideration is limited to a stellar model with the crustal thickness 
$\Delta r/R =0.1$, where $R $ is the radius.
The size depends on the mass or the equation of state of high density-matter.
We also ignore effects of general relativity and temperature.
More realistic treatments lead to some modifications.
The difference is a few in the magnitude, and overall property is unchanged. 
Rather, the ambiguity associated with the strength and configuration of magnetic field is larger than that with these approximations.
We assume that the magnetic field is always expelled from the core, and for a fixed crustal model, focus on the effects of the initial magnetic field on its evolution.
Electron number density $n_{\rm e}$, 
conductivity $\sigma$ in eq.(\ref{Edef.eqn}),
and viscous coefficient $\nu$ in the crust have
the same functional forms given in eqs.(28)--(30) in Paper I. 
For the normalization of time, we use 
$\tau_{\rm H}= 6.9\times 10^{5}(B_{0}/5\times10^{13}{\rm G})^{-1}$yr
\footnote{%
For the discussion, we adopt an intermediate value between a magnetar and a pulsar,
$5\times10^{13}{\rm G}$,
}, which is the Hall timescale calculated by the field configuration, $\Psi_{(1)}$, using
the magnetic field strength, $B_{0}$, at the poles ($r=R,\theta=0$).
The typical Ohmic dissipation time defined by the electronic conductivity at the core--crust interface is 
$\tau_{\rm Ohm}= 5.7\times 10^{6}$yr. 
However, the actual dissipative time is much shorter than $\tau_{\rm Ohm}$ because
the conductivity decreases toward the surface.
The Ohmic time is also proportional to the square of the spatial size, and hence, the local irregularity is expected to be smoothed on a smaller timescale.
The magnetic field evolution for the initial model, $\Psi_{(1)}$, is shown in 
Fig.~\ref{Fig1}.
The left panel displays the evolution of the magnetic energy, $E_{\rm B}$, which is normalized by 
$B_{0}^2R^3=3\times 10^{45}(B_{0}/5\times10^{13}{\rm G})^{2}$erg.
The magnetic energy decreases by half at 
$t\approx 0.3\tau_{\rm H}=2\times 10^{5}(B_{0}/5\times10^{13}{\rm G})^{-1}$yr.
The thick lines represent the results when the bulk motion in the crust is considered.
The magnetic energy is further decreased on being converted to mechanical motion.
The behavior of the time-integrated Joule loss is almost unchanged irrespective of the viscous bulk flow.
We typically examine energy conservation in our numerical calculation.
The magnetic energy loss is the sum of the time-integrated Joule loss and the mechanical work for the bulk motion.
A significant amount of energy is converted into mechanical motion.
For example, although the total work is $0.4 B_{0}^2R^3$ at $ 0.3\tau_{\rm H}$, 
this is not necessarily 
the amount of kinetic energy of the bulk motion at $0.3\tau_{\rm H}$.
In actual cases, the energy is dissipated through bursts, flares, or other processes, which are not treated here.
Regardless of the detailed process, energy is transferred to the bulk motion, and the magnetic energy loss is enhanced.
The exterior of the star is assumed to be vacuum.
The external magnetic field is described by a sum of the multipoles,
\begin{equation}
{\VEC B} =-{\VEC \nabla}\left(\sum \frac{\alpha_{l}}{r^{l+1}} P_{l} \right),
\end{equation}
where $P_{l}(\cos \theta)$'s are Legendre polynomials, and the
multipole moments, $\alpha_{l}$, are calculated by matching to the interior component $B_{r}$.
The time evolution of $\alpha_{l}/(B_{0} R^{l+2})$ $(l=1,3,5)$ is shown in the right panel of Fig.~\ref{Fig1}.
The moments with even number $l$ vanish owing to a symmetry.
The dipole moment slightly decreases and higher moments grow with time.
We also display the magnetic field strength at the poles ($\theta =0, \pi$), i.e.,
$B_{s}= B_{r}(R,0)$.  
Note that for a dipolar field, $ B_{r}(R,0)=-B_{r}(R,\pi)$.
Although both $\alpha_{1}$ and $B_{s}$ decrease with time, they are not equal.
The field strength, $B_{s}$, at the polar caps decreases more rapidly than the dipole moment, $\alpha_{1}$.
This difference can be attributed to $B_{s}$ containing the contribution of higher multipole components, which are produced by the nonlinear coupling in the evolutionary equation: 
$B_{s} \approx 2\alpha_{1}R^{-3}+4\alpha_{3}R^{-5}+6\alpha_{5}R^{-7}$.
In the model including the bulk motion, $B_{s}$ decreases by $\sim 25$\% at 
$t\approx 0.3\tau_{\rm H}$ and $\alpha_{1}$ by $\sim 5$\%.

   \subsection{Model containing poloidal closed field}
\begin{figure}
\begin{center}
  \includegraphics[scale=1.0]{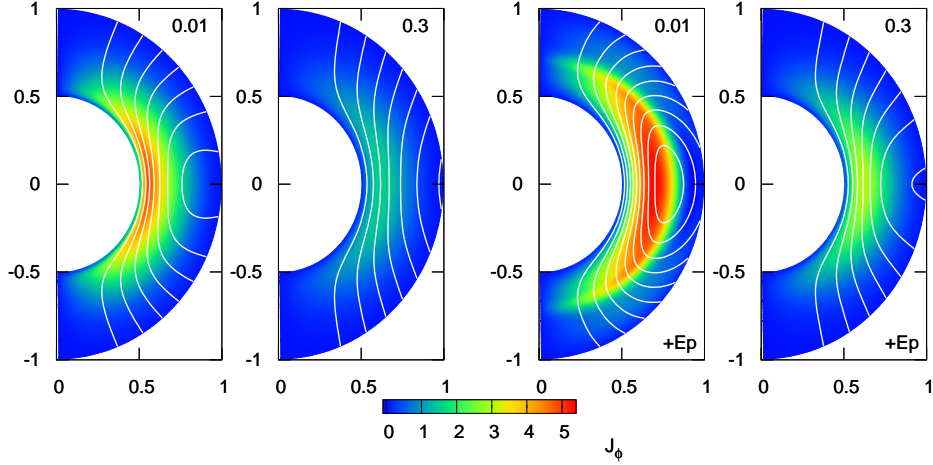}%
\caption{ 
\label{Figj3}
Snapshots of poloidal magnetic function 
$\Psi$ and azimuthal current $j_{\phi}$ at $t=0.01\tau_{\rm H}$ and 
$t=0.3\tau_{\rm H}$.
Contours of magnetic function 
represent $\Psi/(B_{0}R^2) =0.07\times n~(n=1,2,\cdots)$.
Color indicates strength of azimuthal current 
$10^{-2}\times 4\pi j_{\phi}R/(cB_{0})$.
Left two panels show evolution in barotropic MHD equilibrium.
Right two panels present model with extra poloidal magnetic field.
It should be noted that crust region ($0.9\le r/R\le 1$) 
is enlarged five times for display.
}
\end{center}
\end{figure}

We consider another initial model that deviates from the barotropic MHD equilibrium.
Although the magnetic field is purely poloidal, it has a loop structure, as depicted in 
Fig.~\ref{Figj3}.
We add the field, $\Psi_{(2)}$, to $\Psi_{(1)}$, in the fiducial model as follows:
\begin{equation}
\Psi_{(2)}=N_{1}[(r-r_c)(r-R)]^2\sin^2\theta,
\label{eqnaddPsi}
\end{equation}
where $r_c (=0.9R)$ is the core radius and $N_{1}$ is the normalization constant.
The initial current is accordingly changed as 
$j_{\phi {(1)}}+j_{\phi {(2)}}$, where $j_{\phi {(2)}}$ is associated with 
$\Psi_{(2)}$.
The total magnetic energy may increase on changing $N_{1}$.
However, the magnetic field at the surface is unchanged from that of $\Psi_{(1)}$,
because the function (\ref{eqnaddPsi}) and its derivative vanish at the surface. 
We demonstrate the effect of the extra field, $\Psi_{(2)}$,
for which $N_{1}=3.2\times 10^{4}B_{0}R^{-2}$ is used.
The maximum field strength is $\sim 22B_{0}$ due to $\Psi_{(2)}$,
and the maximum  position is $(r, \theta)\approx ((r_{c}+R)/2, \pi/2)$.
Figure~\ref{Figj3} shows the Hall evolution for two initial models: 
$\Psi_{(1)}$ in the left two panels and $\Psi_{(1)}+ \Psi_{(2)}$ in the right two panels.
To display a detailed structure, 
the crustal region, $(r\sin\theta, r\cos\theta)$, ($0.9\le r/R \le 1$) 
is enlarged five times in the figure as 
$(\xi(r)\sin\theta, \xi(r)\cos\theta)$,
where $\xi=1+(R-r)/(2\Delta r)$ and $\Delta r/R=0.1$.
The poloidal magnetic field and the azimuthal current are displayed at their representative times 
$t/\tau_{\rm H}=$0.01 and 0.3.
A large amount of the current is initially stored in the model containing the loop.
By including $j_{\phi (2)}$, the maximum point of $j_{\phi}$ is shifted to the middle of the crust, where $\Psi$ is the largest.
At time $t=0.3\tau_{\rm H}$, the loop structure vanishes, and a large amount of the current is dissipated. 
The magnetic function becomes approximately the same as that in the fiducial model.
The difference between the initial models is not clear at $t=0.3\tau_{\rm H}$.

\begin{figure}
\begin{center}
  \includegraphics[scale=0.80]{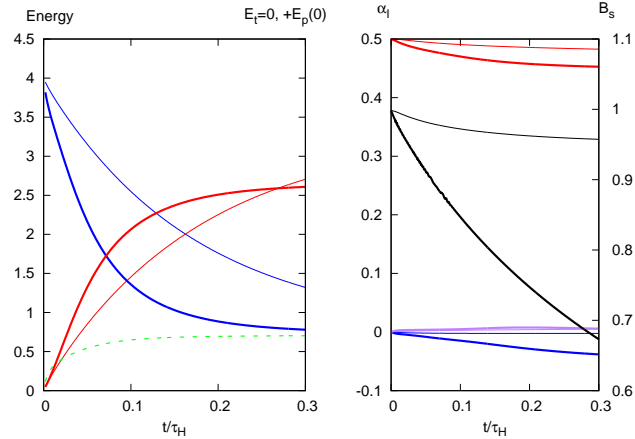}
\caption{ 
 \label{Fig3loop}
Left panel shows decay of magnetic energy of
initial field $\Psi_{(1)}+ \Psi_{(2)}$.
Magnetic energy $E_{\rm B}/(B_{0}^2R^3)$ is plotted as blue decreasing curve,
time-integrated Joule loss as red increasing curve,
and work by green dotted curve.
Right panel displays variation in magnetic multipole moments,
$\alpha_{1}/(B_{0}R^{3})$ (in red color),
$\alpha_{3}/(B_{0}R^{5})$ (in blue color),
$\alpha_{5}/(B_{0}R^{7})$ (in purple color).
Field strength 
$B_{r}/B_{0}$ at polar surface with right scale 
is plotted by black curve.
Thick curves represent model in which bulk motion is considered.
}
\end{center}
\end{figure}

%
Figure~\ref{Fig3loop} shows the evolution of the magnetic energy.
The initial value of $E_{\rm B}/(B_{0}^2R^3)=4$ decays to 
$E_{\rm B}/(B_{0}^2R^3)\sim 1$ at $t=0.3\tau_{\rm H}$.
The energy at $0.3\tau_{\rm H}$ is almost the same as that considered in 
Fig.~\ref{Fig1}.
The decay curve becomes steeper than that for the fiducial model; however, most of the energy dissipated during this period is associated with the initial extra components, 
$\Psi_{(2)}$ and $j_{\phi (2)}$.
The long-lived component is unchanged irrespective of the initial models,
 $\Psi_{(1)}$ or $\Psi_{(1)}+\Psi_{(2)}$,
as inferred from Fig.~\ref {Figj3}.
In the right panel of Fig.~\ref{Fig3loop}, the evolution of the field strength, 
$B_{s}$, at the polar caps and the dipole moment, $\alpha_{1}$, are shown.
The poloidal field initially confined in the interior slightly 
enhances their decrease.

   \subsection{Model containing large toroidal field}
\begin{figure}
\begin{center}
  \includegraphics[scale=0.8]{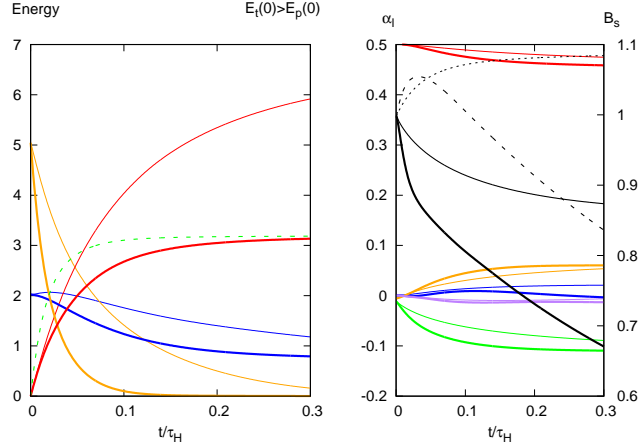}%
\caption{ 
 \label{Fig4tor}
Evolution of magnetic field for initially toroidal-field-dominated model.
The left panel shows magnetic energies of poloidal component (in blue color)
and toroidal component (in orange color).
Time-integrated Joule loss is represented by curve (in red color).
Dotted line (in green color) denotes mechanical work for 
model with bulk flow.
Right panel displays time-variation in magnetic multipole moments,
$\alpha_{1}/(B_{0}R^{3})$ (in red color),
$\alpha_{2}/(B_{0}R^{4})$ (in green color),
$\alpha_{3}/(B_{0}R^{5})$ (in blue color),
$\alpha_{4}/(B_{0}R^{5})$ (in orange color),
$\alpha_{5}/(B_{0}R^{7})$ (in purple color).
Field strength $B_{r}(R,0)/B_{0}$ with right scale 
is plotted by black line and 
$-B_{r}(R,\pi)/B_{0}$ by black dotted line.
Thin line represents result for purely Hall evolution,
whereas thick line result with bulk flow.
}
\end{center}
\end{figure}

\begin{figure}
\begin{center}
  \includegraphics[scale=1.5]{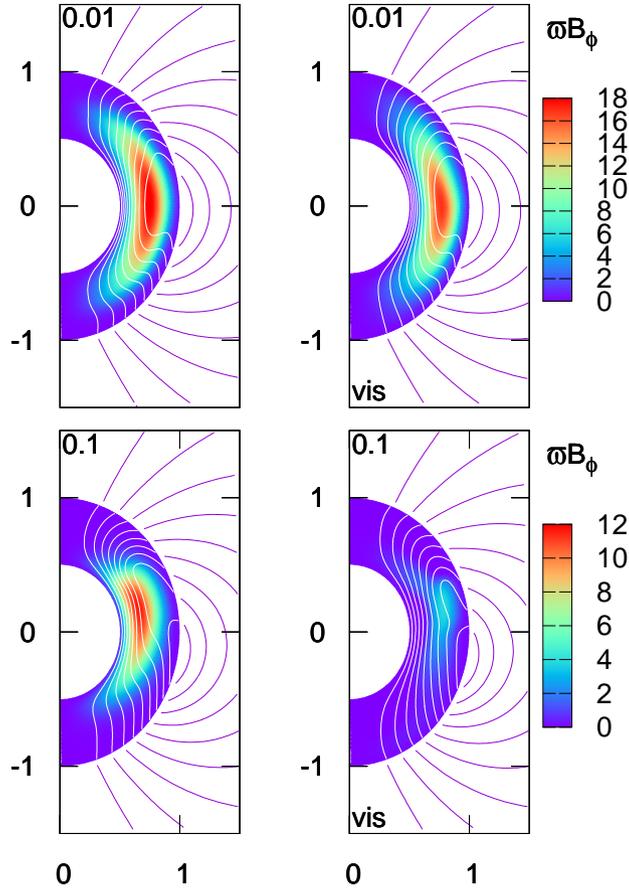}%
\caption{ 
\label{Figgs22}
Snapshots of poloidal magnetic function
$\Psi/(B_{0}R^2)$ by contour lines 
and current function $ \varpi B_{\phi}/(RB_{0})$ in color
at $t=0.01\tau_{H}$(top) and $t=0.1\tau_{H}$(bottom).
Left panels show results of purely Hall evolution, and 
right panels those of models including bulk motion.
}
\end{center}
\end{figure}

The initial model considered in the previous subsections considered a purely poloidal magnetic field.
Generally, a toroidal component is produced during the magnetic field evolution; however, it is significantly weaker than the poloidal one.
The maximum of the toroidal field is one order of magnitude 
smaller than that of the poloidal field. 
The angular dependence of magnetic field 
near the surface is typically approximated by the dipolar field.
It is important to study how these behaviors may
change with the initial toroidal magnetic field.
However, it is difficult to construct toroidal-field-dominated models 
that satisfy the MHD equilibrium condition.
To examine the effect of a large toroidal field confined to a crust, we add
the following toroidal component to the fiducial model described in Section 3.1:
\begin{equation}
S=(r \sin \theta)B_{\phi} =N_{2}[(r-r_c)(r-R)]^2\sin^2\theta,
\end{equation}
where $N_{2}$ is the normalization constant.
By increasing $N_{2}$, we construct models with a strong toroidal field.
However, the state does not satisfy the barotropic MHD equilibrium
condition (see eq.(\ref{MHDeqil.eqn})).
Specifically, the poloidal field,
 $\Psi_{(1)}$, is not a solution for eq.~(\ref{MHDeqil.eqn}) and 
$S \ne S(\Psi_{(1)})$. 
Figure \ref{Fig4tor} shows the evolution of the model in which the toroidal magnetic energy is initially dominated.
We use $N_{2}=3.1 \times 10^2 B_{0}R^{-3}$, and the maximum of $B_{\phi}$ is 
$\sim 20 B_{0}$, which is comparable to that of poloidal component. 
The magnetic energy ratio of the toroidal and poloidal components is initially 
$E_{t}/E_{p}=2.5$.
The time evolutions of $E_{p}$ and $E_{t}$ are shown in the left panel of Fig.~\ref{Fig4tor}.
The energy transfer from the toroidal to poloidal components occurs only in the early times at $\approx 0.01\tau_{\rm H}$.
The toroidal energy decays extremely rapidly; therefore, the increase in the poloidal energy, which is transferred during the early times, is small.
Thus, the evolution of the poloidal energy is almost the same as that in Fig.~\ref{Fig1}. 
For the model considering the bulk motion, the increase is insignificant because the energy is also transferred to another channel, i.e., the bulk flow motion.
The initial toroidal magnetic energy is converted either to Joule loss or mechanical work. 
Consequently, the toroidal component rapidly decays; thus, it does not significantly affect the poloidal field evolution.
The right panel of Fig.~\ref{Fig4tor} shows the external 
multipole moments, $\alpha_{l}/(B_{0}R^{l+2})$,
and the field strength, $B_{s}$, at the polar caps.
Under the initial toroidal magnetic field, multipole components with even number $l$ are produced, so that the relation, 
$B_{r}(R,0)=-B_{r}(R,\pi)$, is broken, unlike under the dipolar field ($l=1$).
Subsequently, we will discuss the asymmetry of the magnetic field based on 
Fig.~\ref{Figgs22}.
Both $B_{r}(R,0)$ and $-B_{r}(R,\pi)$ are displayed as two curves in 
Fig.~\ref{Fig4tor}.
The evolution of the averaged value,
$(|B_{r}(R,0)|+|B_{r}(R,\pi)|)/2$, is similar to the curve in the right panel of Fig.~\ref{Fig1}.
Thus, we find that although the initial toroidal field causes the asymmetry at the surface, it does not change the average strength of the poloidal field.

\begin{figure}
\begin{center}
  \includegraphics[scale=1.5]{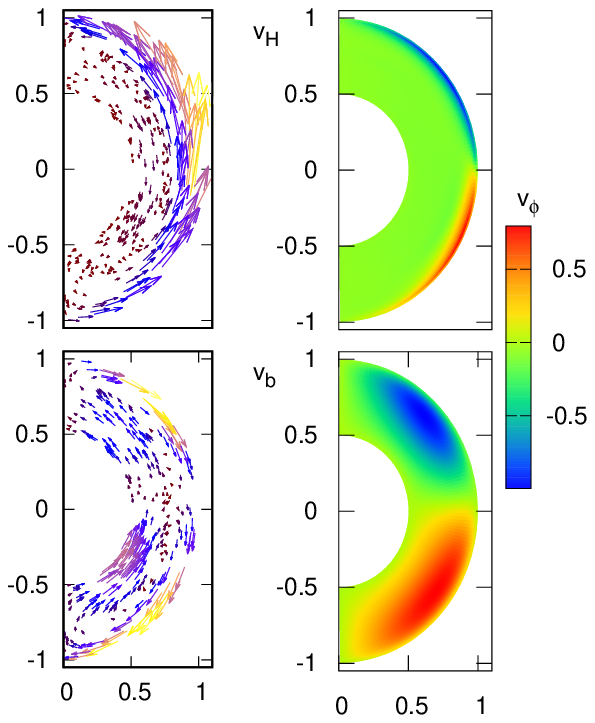}%
\caption{ 
\label{Fgxxx}
Snapshot of velocity fields ${\VEC v}_{\rm H}$ (top) and ${\VEC v}_{\rm b}$ (bottom).
Left panels show poloidal components at randomly selected grid points, and right panels show normalized azimuthal-component by color contour.
}
\end{center}
\end{figure}

%
We demonstrate the magnetic configurations at representative times 
$t/\tau_{\rm H}=$0.01 and 0.1.
Figure~\ref{Figgs22} shows the poloidal magnetic function, $\Psi$, as contour lines and the toroidal field, $\varpi B_{\phi}$, in color in an enlarged coordinate, 
$(\xi(r)\sin\theta, \xi(r)\cos\theta)$.
The poloidal field in the interior is significantly affected by the strong toroidal component.
However, the exterior field is described well by the dipolar one.
In the detailed structure, there is 
asymmetry between the north and south hemispheres in the bottom panels 
($t=0.1\tau_{\rm H}$).
This leads to the difference in the field strengths at the two poles, as shown in the right panels of Fig.~\ref{Fig4tor}.
The toroidal component rapidly vanishes in the model considering the bulk motion, as shown in the right panels.
At time $t/\tau_{\rm H} =0.1$, the maximum of $B_{\phi}$ decreases to up to 30\% of the initial value.
The toroidal component does not affect the subsequent evolution once
its energy becomes less than that of the poloidal one.
Figure \ref{Fgxxx} shows behaviors of the Hall velocity
${\VEC v}_{\rm H}\equiv -{\VEC j}/(en_{\rm e})$ and the bulk velocity ${\VEC v}_{\rm b}$
at $t/\tau_{\rm H}=0.01$.
The snapshot of the magnetic field corresponds to the right top panel of 
Fig.~\ref{Figgs22}.
Both velocities are normalized by a maximum value 
$\sim 30 R/\tau_{\rm H}$$\sim 40$cm yr$^{-1}$.
The tangential components $ v_\theta$ and $v_\phi$ are generally larger than radial component: typically $v_r \times R/\Delta r \sim v_\theta \sim v_\phi$.
The left top panel of Fig.~\ref{Fgxxx} shows the poloidal component of the Hall velocity.
The initial toroidal field $B_{\phi}>0$ induces a global circulation of poloidal current, which clockwise flows in $r-\theta$ plane.
The direction of the Hall velocity is therefore opposite, 
and the magnitude becomes small in a deeper region owing to 
$|v_{\rm H}|\propto 1/n_{\rm e}$.
The azimuthal component shown in Fig.~\ref{Fgxxx} is also large near the surface, although the current flows in entire crust.
In the bottom panels of Fig.~\ref{Fgxxx}, the bulk flow velocity is shown.
The angular dependence of poloidal velocity-field is approximately dipolar, 
whereas that of the azimuthal field is quadrupolar.
These flow patterns come from the initial toroidal field $B_{\phi}$.
The flow velocity ${\VEC v}_{\rm b}$ without it was shown in Fig.4. of Paper I:
quadrupolar in poloidal component  ${\VEC v}_{{\rm b} p}$ and
dipolar in azimuthal one $v_{{\rm b} \phi}$.

\section{Evolution of quadrupole deformation}
In this section, we examine the effect of bulk motion on the magnetic deformation, which should evolve with the magnetic field.

\subsection{Ellipticity}
The irrotational part of the Lorentz force is related to the deviation from a spherical shape. 
We solve eq.(\ref{eqndf2}) for $\delta \chi$ inside the crust of a neutron star.
We assume that the magnetic field does not penetrate 
the core always.
Because the core is spherically symmetric, 
the boundary condition at the core--crust interface, $r=r_{c}$,
is $\delta \chi=0$.
At the surface, $r=R$, the condition is 
$\partial \delta \chi/\partial r=0$, so that $\delta f_{r}=0$ in eq.
(\ref{eqndecomp}).
The magnetic deformation in the crust can be estimated using the potential,
 $\delta \chi$, which may be incorporated as an additional pressure term in eq.(\ref{Forcebalance.eqn}).
The change in an isobaric surface is expressed by the displacement, $\xi(r, \theta)$, 
which is 
\begin{equation}
\xi=- \frac{\delta \chi}{dp/dr}=
\frac{\delta \chi r^2}{G\rho M_{r}},
   \label{radialdisp.eqn}
\end{equation}
where a hydrostatic condition (\ref{eqnbackground})
is used in the last expression.
Using $\xi$, we estimate the quadrupole deformation of the crustal shape as 
\begin{equation}
\epsilon_{\rm S}= \frac{2\xi(r, \pi/2)-(\xi(r, 0)+\xi(r, \pi))}{2r}.
\label{Defqshape}
\end{equation}
Shape deformation,
$\epsilon_{\rm S}$ depends on the radius; therefore, hereafter, we utilize 
$\bar{\epsilon}_{\rm S}$, the average value in the crust.
\begin{equation}
\bar{\epsilon}_{\rm S}=
10^{-6}\epsilon_{*}\left(
\frac{B_{0}}{5\times 10^{13}{\rm G}}\right)^2,
%
\end{equation}
where we use the canonical values for the mass 
($M=1.4 M_\odot$) and radius ($R=12{\rm km}$) of a neutron star. 
The coefficients, $\epsilon_{*}$, are numerically calculated for the evolving models.
The quadrupole deformation 
is important in spin-down evolution of a pulsar, 
and hence, also for gravitational wave astronomy
\citep[e.g.][]{2017ApJ...844..112G,2019MNRAS.490.2692K}.
The ellipticity, $\epsilon_{\rm Q}$,
is defined by the quadrupole moment as follows: 
\begin{equation}
\epsilon_{\rm Q} \equiv \frac{I_{xx}-I_{zz}}{{\bar I}},
\label{Defqmoment}
\end{equation}
where $I_{ij}= \int \rho x_{i}x_{j} d^3x$ and 
${\bar I}$ is the average of $I_{ij}$.
The ellipticity, $\epsilon_{\rm Q}$ depends on the entire stellar structure.
Our consideration of the deformation 
is limited to the crustal part of a star, which leads to
the difference in the numerator of eq.(\ref{Defqmoment}).
The typical mass (or mass density) relevant to the numerator is the crustal mass,
$\Delta M$, whereas that for the denominator is the stellar mass, $M$.
The ellipticity, $\epsilon_{\rm Q}$, is estimated as
\begin{equation}
\epsilon_{\rm Q} \approx \frac{\Delta M}{M}\bar{\epsilon}_{\rm S}
 \approx 10^{-2}\times\bar{\epsilon}_{\rm S},
\label{QdeformI}
\end{equation}
where $\Delta M$ is $\sim 10^{-2}M$ for a typical model.
Note that an exact value of $\epsilon_{\rm Q}$ is not obtained without a specified stellar model.

\begin{figure}
\begin{center}
  \includegraphics[scale=0.80]{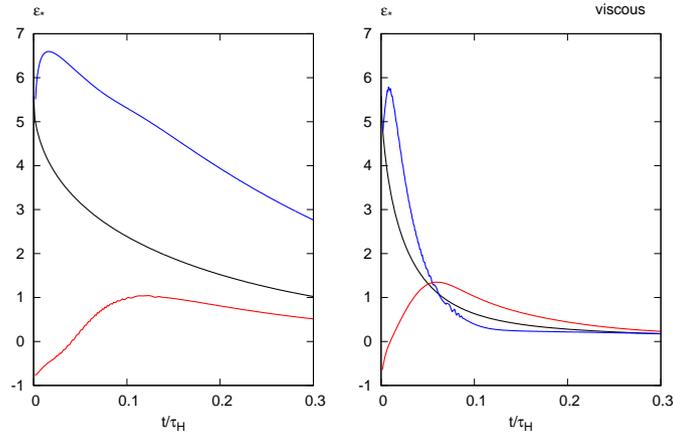}%
\caption{ 
\label{FigDeform}
Evolution of magnetic deformation.
Central curve in black color represents fiducial model described in Section 3.1.
Model initially containing a loop, as discussed in Section 3.2, is represented 
by blue curve, and toroidal-field-dominated model described in Section 3.3
is denoted by a red curve.
In latter case, $\epsilon_{*}$ is initially negative. 
Left panel shows results for Hall evolution, and right one for those with bulk motion.
}
\end{center}
\end{figure}

%
We discuss the evolution of the magnetic deformation of a neutron star.
Figure~\ref{FigDeform} shows the normalized ellipticity, 
$\epsilon_{*}~(\propto \bar{\epsilon}_{\rm S})$, numerically 
calculated for our crustal model. 
It is well known that a poloidal magnetic field produces an oblate shape 
($\bar{\epsilon}_{\rm S} >0$), whereas a toroidal field a prolate one 
($\bar{\epsilon}_{\rm S} <0$).
For the fiducial model, as described in Section 3.1, $\epsilon_{*}$ is always positive and
monotonically decreases with time.
The value at $t/\tau_{\rm H}=0.3$ decreases to $\sim$ 20\% of that with the initial model.
This decrease in $\epsilon_{*}$ originates from the 
decrease in the magnetic energy or magnetic pressure.
For the model containing a loop, as discussed in Section 3.2,
the Lorentz force from the extra part initially added by
$\Psi_{(2)}$ is canceled  
when calculating the quadrupole deformation throughout the crust. 
According to the prompt redistribution of the current, the 
deformation increases at $t/\tau_{\rm H}\sim 0.01$.
With further loss of the magnetic energy, the deformation gradually decreases.
The initial value of $\epsilon_{*}$ is negative for the toroidal-field-dominated model described in Section 3.3.
The associated prolate shape changes to an oblate one based on the magnetic field evolution.
The change in the sign of $\epsilon_{*}$ broadly corresponds to the turning point of the dominant component. 
The decay timescales of the poloidal and toroidal fields are different.
The toroidal field decays more rapidly than the poloidal one, which 
causes the change from the prolate to an oblate shape on a secular timescale.
In the right panel of Fig.~\ref{FigDeform}, we show the evolution of the magnetic deformation for the models considering a viscous bulk motion.
As presented in Figs.~\ref{Fig1}, \ref{Fig3loop}, and \ref{Fig4tor}, the
magnetic energy is converted into the fluid kinetic energy and the 
Joule heat, which are of the same order.
The decay timescale becomes approximately half.
This rapid dissipation of the magnetic energy shortens the evolutionary
timescale of $\epsilon_{*}$, as shown in Fig.~\ref{FigDeform}.

 \subsection{Discussion}
\citet{2016MNRAS.459.3407S}
already considered the evolution of the quadrupole ellipticity caused by the Hall drift in a neutron star crust.
Our methods and models significantly differ from those used by them; therefore, a 
comparison of the numerical results is difficult.
Their evolution model is based on a relatively detailed magneto-thermal calculation.
The most crucial difference is with their mathematical method for evaluating the ellipticity.
They calculated the quadrupole deviation based on a nonbarotropic nature.
Specifically, the quadruple moment, 
$\delta I_{ij}=\int \delta \rho x_{i}x_{j}d^3x$, was calculated for a
nonspherical perturbation $\delta \rho$ of the density
induced by a 'curl' of the Lorentz force
\citep[][]{2013MNRAS.434.1658M,2015MNRAS.447.3475M}.
This corresponds to the solenoidal part of the Lorentz force and 
induces a circulation of the viscous fluid, as discussed in Section 2.
In our treatment, the quadrupole deviation is calculated 
from the irrotational part of the force.
When we apply their method to our initial model in the barotropic 
MHD equilibrium, the ellipticity is exactly zero.
However, our method yields 
$\epsilon_{\rm Q}\approx10^{-8}(B_{0}/5\times 10^{13}{\rm G})^2$ in
 eq.(\ref{QdeformI}).
Therefore, two methods generate significantly different ellipticities.
Moreover, the time evolution of the deformation in their models is complex.
No clear relationships between the ellipticity and the maximum amplitude of the poloidal or toroidal magnetic field are revealed in their results.
In the present treatment, the quadrupole deformation, $\epsilon_{\rm Q}$, gradually decreases with the magnetic energy, $E_{\rm B}$,
although $\epsilon_{\rm Q}$ is not exactly proportional to $E_{\rm B}$.
The oblate or prolate shape of the deformation depends on whether the dominant magnetic field is poloidal or toroidal.
The ellipticity, $\epsilon_{\rm Q}$ in eq.(\ref{QdeformI})
coincides with the energy ratio, $E_{\rm B}/E_{\rm G}$, where
the magnetic energy stored in the crust is
$E_{\rm B}\approx 6\times 10^{45}(B_{0}/5\times10^{13}{\rm G})^2$erg
and the gravitational binding energy of the star is $E_{\rm G} \approx 3\times 10^{53}$erg.
This relation between the quadrupole deformation and the energy ratio
has been confirmed by various models,
irrespective of the equation of states and the relativistic factor,
and presented in the literature
\citep[e.g.,][]{1954ApJ...119..407F,1996A&A...312..675B,
1999A&A...352..211K,2001MNRAS.327..639I,2006ApJS..164..156Y,
2008MNRAS.385..531H,2009MNRAS.395.2162L,2014PhRvD..90j1501U,
2019PhRvD.100l3019U}.
In the previous models, the magnetic field extends into the whole interior 
\footnote{%
The ellipticity depends on the field strength as
$\epsilon_{\rm Q} \propto B^2$, but it 
changes as $\epsilon_{\rm Q} \propto BH_{c}$
for a neutron star containing a type II superconducting core, where
$H_c$ is the critical field\citep{2013PhRvL.110g1101L}
}. However,
it is localized to the outer part in our model.
We found that the empirical relation between quadrupole
deformation and energy ratio is still applicable. 

\section{Concluding remarks}
In this study, we consider
the effects of bulk motion on the magnetic field evolution in a neutron star crust.
Particularly, the decay of both the surface dipole and the magnetic deformations
are explored.
Their evolution is always accelerated by the energy loss 
to the bulk flow as a new channel.
Consequently, the deformation in the crust shrinks with decreasing 
magnetic energy.  
The toroidal and higher poloidal components 
decay on a smaller timescale than that of the dipolar component.
Finally, the quadrupole deformation is determined by the long-lived dipole. 
From our numerical simulation, we also find that the deformation is 
a good indicator of magnetic energy stored in the interior.
The shape is prolate when the toroidal component is initially
dominated, whereas it becomes oblate when the poloidal component is dominated.
The ratio of the magnetic energy to the gravitational binding energy 
provides the ellipticity.
Specifically, the magnetic energy of the dipole component ($l=1$)
is important because the quadrupole deformation, which
corresponds to the spherical harmonics index, $l=2$,
is related to the magnetic field at $l=1$.
The contribution from higher order components with $l>1$ is partially
canceled by the spatial integration.
The magnetic field evolution depends on the initial configurations, which are
less clear.
The dynamical stability of a magnetized star is a long-standing unsolved 
problem\citep[e.g.,][]{2006A&A...450.1077B,2012MNRAS.424..482L,
2013MNRAS.435L..43C,2020MNRAS.495.1360S}.
Some MHD calculations suggest stable models of twisted torus,
a configuration comprised of poloidal and toroidal components.
They are based on a whole star without a crust, and the formation and 
stability of the localized fields is unclear.
As plausible states at the birth of a magnetized star, we
consider a barotropic MHD equilibrium model and its variants.
The crustal magnetic field, which 
initially satisfies the MHD equilibrium, evolves toward the Hall equilibrium.
The difference between the two equilibria originates from the different 
distributions of the mass and electron number densities\citep{2013MNRAS.434.2480G,2014MNRAS.445.2777F}.
The spatial profile is not significantly different; therefore, the transition from the barotropic MHD to the Hall equilibria is not extremely drastic.
To examine the effect of the initial field configuration, 
we add poloidal and toroidal fields as extra components and 
compare the evolution.
Our initial model presented in this paper 
is an ad hoc method to demonstrate the difference on a secular timescale.
Owing to the extra poloidal or toroidal components, different evolutionary features are expected; however, our results show that the extra components vanish and do not have a significant effect. 
We find that the surface dipole field 
given by the barotropic MHD equilibrium is almost stable, not changing significantly in $\sim 10^6$ years.
A counter example can be easily considered.
Let us suppose that the electric current relevant to the dipole field is localized near the surface where the electric conductivity is low.
For this case, the bulk motion is also effective in that region owing to the low viscosity there
\citep[][]{2019MNRAS.486.4130L,2020MNRAS.494.3790K}.
The decay timescale correspondingly becomes shorter, even 
less than $10^5$ yr for the extreme configurations.
In summary, evolutionary models for a simple magnetic field
are obtained under the assumption that 
the initial model comprises ordered smooth fields in near-MHD equilibrium.
By relaxing the above assumption, we expect that a rapid change in the locally irregular components will lead to diverse evolutionary behaviors.

 \section*{Acknowledgements}
%
 This work was supported by JSPS KAKENHI Grant Number 
JP17H06361, JP19K03850(YK), JP18H01245, JP18H01246, JP19K14712 (SK), 
JP20H04728 (KF).
%

 \section*{DATA AVAILABILITY}
%
The data underlying this article will be shared on reasonable request 
to the corresponding author.
%

 \bibliographystyle{mnras}
 \bibliography{kojima20Oct} 

\end{document}